\documentclass[12pt]{iopart}

\usepackage{iopams}
\usepackage[utf8]{inputenc}
\usepackage[english]{babel}
\usepackage{caption}
\usepackage{subcaption}
\usepackage[table,xcdraw,usenames,dvipsnames]{xcolor}
\usepackage{placeins}
\usepackage{graphicx}
\usepackage{comment}
\usepackage[colorlinks=true, pdfborder={0 0 0}, citecolor=blue,linkcolor=red]{hyperref}
\usepackage{pdfsync}
\usepackage{cite}

\usepackage{tikz,pgfplots}
\usetikzlibrary{arrows}
\usetikzlibrary{decorations.markings}
\usetikzlibrary{math}
\usetikzlibrary{positioning,calc}
\usetikzlibrary{backgrounds}
\newcommand{\dif}{\text{d}}

\newcommand{\ronum}[1]{ \textup{\uppercase\expandafter{\romannumeral#1}}}

\makeatletter
\DeclareRobustCommand{\text}{%
  \ifmmode\expandafter\text@\else\expandafter\mbox\fi}
\let\nfss@text\text
\def\text@#1{{\mathchoice
  {\textdef@\displaystyle\f@size{#1}}%
  {\textdef@\textstyle\f@size{#1}}%
  {\textdef@\textstyle\sf@size{#1}}%
  {\textdef@\textstyle \ssf@size{#1}}%
  \check@mathfonts
  }%
}
\def\textdef@#1#2#3{\hbox{{%
                    \everymath{#1}%
                    \let\f@size#2\selectfont
                    #3}}}
\makeatother

\begin{document}

\title[Phase shift in periodically driven non-equilibrium systems]{Phase shift
in periodically driven non-equilibrium systems: Its identification and a bound}
\author{Julius Degünther, Timur Koyuk and Udo Seifert}
\address{II. Institut für Theoretische Physik, Universität Stuttgart, 70550 Stuttgart, 
Germany}
\ead{deguenther@theo2.physik.uni-stuttgart.de}
\begin{abstract}
Time-dependently driven stochastic systems form a vast and manifold class of
non-equilibrium systems used to model important applications on small length
scales such as bit erasure protocols or microscopic heat engines. One property
that unites all these quite different systems is some form of lag between the
driving of the system and its response. For periodic steady states, we quantify this lag by introducing a generalized phase difference and prove a
tight upper bound for it. In its most general version, this bound depends only on the
relative speed of the driving.
\end{abstract}


\begingroup
\let\newpage\relax
\maketitle
\endgroup

\section{Introduction}
Recent technological advancements allowing a precise manipulation and
fabrication on the micro- and nanoscale led to a growing interest in small
stochastic non-equilibrium systems in which fluctuations play a fundamental
role. Time-dependently driven systems constitute the arguably largest and most
diverse class of such non-equilibrium systems containing a variety of
systems such as microscopic heat
engines~\cite{schm08, espo10, blic12, abah12, zhan14, bran15b, camp16a}, applications of bit erasure
processes~\cite{beru12,proe20,proe20a,zhen21}, stochastic pumps \cite{raha08, cher08, asba15, raz16a, rots16}, as well as
biological or chemical systems exposed to time-dependent
stimuli, see, e.g., Refs.~\cite{haya12a, erba15}. One common property of such
time-dependently driven systems is some form of lag between the driving and the
response of the system. Time-dependent driving is generally accompanied by a
change in the instantaneous steady state to which the system tends to relax.
The relaxation is not immediate and inevitably leads to a lag or
delay between the system and its driving. Previously, this lag has been
quantified by comparing the time-dependent probability density with the instantaneous steady
state leading to relations about
dissipated work and drift~\cite{vaik09, frez17}.

Exact solutions for non-equilibrium systems are rarely feasible, even less so
for time-dependently driven systems. However, major progress has been achieved
by establishing inequalities and bounds for various dynamic quantities. One
vast family of bounds are the so-called thermodynamic uncertainty relations \cite{horo20},
which cover a wide range of different system classes such as non-equilibrium
steady states~\cite{bara15, ging16}, systems relaxing to equilibrium \cite{dech17, liu19}, periodically driven
systems~\cite{proe17, bara18b, koyu19a, bara18c} or arbitrarily driven
systems~\cite{proe19, koyu20}. Besides the thermodynamic uncertainty relations
speed limits~\cite{shir18, shir19, ito20, vo20, yosh21} form another large group of
bounds. The general idea of these speed limits is often related to the
aforementioned notion of lag since there are inherent limitations for the speed of
certain processes due to "inertia" of these systems.

It is often desirable to minimize the lag in time-dependently driven
systems. For example, biological systems need to readily adapt to changing
circumstances. A fast response to changing concentrations of nutrients can
pose a significant advantage for the organism~\cite{stoc08}. Another example
can be found among measuring devices, such as molecular sensors. For many
applications they need to operate with small lag~\cite{zhan21}.

In this paper, we address the question of quantifying the lag in periodically
driven systems. As a first main result, we introduce a quantity that measures
the lag in a Markov network, for which the energy of each state varies
periodically. In contrast to the approach in Refs.~\cite{vaik09, frez17} this quantity is not
time-dependent but rather poses a phase difference between the driving and the
response of the system. As the second main result of this paper, we will prove
a universal upper bound on this phase that only depends on the speed at which the
system is driven.

\section{Setup}
We consider a discrete Markovian system with $N$
states. The time evolution of the probability $p_i(t)$ to find the system in state
$i$ at time $t$ is governed by the master equation
\begin{equation} \label{eq:me} \partial_tp_i(t)=-\sum_{j\neq
i}\left[p_i(t)k_{ij}(t)-p_j(t)k_{ji}(t)\right],
\end{equation} where $k_{ij}(t)$ denotes the transition rate between states
$i$ and $j$.
The transition rates must fulfill the local detailed balance relation in order to model
the system in a thermodynamically consistent way. They are parametrized as
\begin{eqnarray}
  k_{ij}(t)&=k_{0,ij}e^{\alpha_{ij}\Delta E_{ij}(t)}, \\  
  k_{ji}(t)&=k_{0,ij}e^{-(1-\alpha_{ij})\Delta E_{ij}(t)}
\end{eqnarray} with time-dependent energy difference $\Delta E_{ij}(t)\equiv
E_i(t)-E_j(t)$. The parameter $\alpha_{ij}\in\left[0,1\right]$ determines its
splitting in forward and backward rate. The rate amplitude $k_{0,ij}$ sets the time scale of the
transition. We set $k_BT=1$ throughout this paper. The system is driven by periodically
varying the energies
\begin{equation} \label{eq:general_case} E_i(t) =
E_i^0+E\sum_n\left[c^n_i\sin\left(n\Omega t\right)+d^n_i\cos\left(n\Omega t\right)\right]
\end{equation} for each state $i$ with the frequency
$\Omega=2\pi/\mathcal{T}$. The energy amplitude $E$ sets the general scale of
the driving and allows us to keep the Fourier coefficients $\left\{c_i^n,
d_i^n\right\}$ of the order of $1$. The parameters $E_i^0$ are the constant energy
offsets. In the long-time limit, the system will reach the periodic steady state $p_i^\text{ps}(t)=p_i^\text{ps}(t+\mathcal{T})$. In the following, we will focus on this periodic steady state and omit the superscript $p_i(t)\equiv p_i^\text{ps}(t)$.

The framework of stochastic thermodynamics allows us to identify thermodynamic
quantities, such as heat, work and entropy production~\cite{seif12}. Since we consider the
periodic steady state, we focus on the rates of these quantities averaged over
one period. The rate of work, i.e. power, applied to state $i$ is defined as
\begin{equation} \label{eq:powers}
P_i\equiv\frac{1}{\mathcal{T}}\int_0^\mathcal{T}\dif
tp_i(t)\partial_tE_i(t).
\end{equation} The total entropy production rate in these systems
is identical to the medium entropy production rate, since the stochastic entropy
production vanishes in the average over one period, which leaves us with
\begin{equation} \label{eq:entropy}
\sigma\equiv\frac{1}{\mathcal{T}}\int_0^\mathcal{T}\dif
t\sum_{i,j}p_i(t)k_{ij}(t)\ln\left(\frac{k_{ij}(t)}{k_{ji}(t)}\right).
\end{equation} The entropy
production rate is equal to the heat flux in the system and due to
conservation of energy we have
\begin{equation}
\sigma=\sum_{i=1}^NP_i.
\end{equation}

\section{Definition of the phase for a two-state system}
\label{sec:definition} First, we identify the phase that quantifies the lag
in a two-state system. In this case there is only one timescale $k_0$ for the transitions and one energy difference $\Delta E(t)$, which determines the driving. To motivate the definition of the phase, we first consider a
general Fourier series with coefficients
$\left\{f_n,g_n\right\}$. Such a Fourier series can
also be expressed in terms of amplitudes $\left\{A_n\right\}$ and phases $\left\{\Phi_n\right\}$
\begin{equation} \sum_nf_n\sin\left(n\Omega t\right)+g_n\cos\left(n\Omega
t\right)=\sum_nA_n\sin\left(n\Omega t +\Phi_n\right)
\end{equation} with
\begin{eqnarray} \label{eq:1} \Phi_n &= \left\{\begin{array}{lr}
\arctan\left(g_n/f_n\right), & \text{for } f_n\geq 0 \\
\arctan\left(g_n/f_n\right)-\pi, & \text{for } f_n<0
        \end{array}\right..
\end{eqnarray} The phase $\Phi_n$ follows from the coefficients of the
Fourier series and denotes the phase difference between the terms of the Fourier series and, in
this case, $\sin\left(n\Omega t\right)$.

Similarly, the idea behind our definition of the phase that quantifies the lag
in the periodic steady state of a two-state system
is to expand the probability $p_1(t)$ into a Fourier-like series
\begin{equation} \label{eq:series} p_1(t)=\sum_na_nv_n(t)+b_nw_n(t)
\end{equation} with basis functions $\left\{v_n(t), w_n(t)\right\}$ and
coefficients $\left\{a_n,b_n\right\}$. The two crucial basis functions are
$v_1(t)$ and $w_1(t)$, which we choose as functions of the driving
\eref{eq:general_case}. The ratio of the corresponding coefficients defines a
phase analogous to equation~\eref{eq:1}. This phase can be interpreted as a phase difference between the
driving and the response of the system expressed in terms of $p_1(t)$. All
further basis vectors appearing in the series~\eref{eq:series} complement
$v_1(t)$ and $w_1(t)$ so that $\left\{v_n(t), w_n(t)\right\}$ defines a
complete and orthogonal basis for functions with periodicity $\mathcal{T}$.

We now need to find suitable choices for the functions $v_1(t)$ and
$w_1(t)$. Since the phase is supposed to quantify a lag, slow driving
$\Omega/k_0\to0$ is expected to minimize this quantity and fast driving
$k_0/\Omega\to0$ is expected to maximize it. To fulfill these criteria, the coefficients $b_1$ and $a_1$ need to vanish in the limit of slow driving and in the limit of fast
driving, respectively. We start with the first case. In the quasistatic limit $p_1(t)$
converges to the instantaneous equilibrium distribution
\begin{equation} p_1(t)=p_1^{\text{qs}}(t)=\frac{1}{1+e^{\Delta E(t)}}
\end{equation} with $\Delta E(t)\equiv\Delta E_{12}(t)$. The condition that the coefficient $b_1$ vanishes for slow driving is
equivalent to $w_1(t)$ being orthogonal to $p_1^{\text{qs}}(t)$
\begin{equation} \label{eq:a_integral_1}
\int_0^\mathcal{T}p_1^{\text{qs}}(t)w_1(t)\dif
t=\int_0^\mathcal{T}\frac{1}{1+e^{\Delta E(t)}}w_1(t)\dif t\stackrel{!}{=}0.
\end{equation} This condition suggests the choice
$w_1(t)\equiv\partial_t\Delta E(t)/N_2$ with an arbitrary constant $N_2$. The integral
\eref{eq:a_integral_1} becomes zero with this choice, since the integrand is a
function of the periodic $\Delta E(t)$ multiplied with its derivative
$\partial_t\Delta E(t)$. Any integral over one period $\mathcal{T}$ of an integrand
$f\left(g(t)\right)\partial_t{g}(t)$ with periodic function $g(t)=g(t+\mathcal{T})$ vanishes
as shown through a substitution $t\to u=g(t)$. With this choice the
integral~\eref{eq:a_integral_1} becomes proportional to the entropy production
$\sigma$, which generically vanishes in the limit of
slow driving. Thus, we have found the first basis function $w_1(t)$.

Finding $v_1(t)$ is slightly more involved. The criterion for $a_1$ to vanish
in the limit of fast driving translates into the condition that $v_1(t)$ is
orthogonal to $p_1(t)$ in the limit $k_0/\Omega\to 0$. To calculate the
leading orders of $p_1(t)$ in $k_0/\Omega$, we use a perturbative expansion of
the probability
\begin{equation}\label{eq:fast_pert_series}
p_1(\tau)=p_1^{(0)}(\tau)+\frac{k_0}{\Omega}p_1^{(1)}(\tau)+\mathcal{O}\left(\frac{k_0^2}{\Omega^2}\right)
\end{equation} with dimensionless time variable $\tau\equiv\Omega t$. Inserting
\eref{eq:fast_pert_series} into the master equation~\eref{eq:me} yields
\begin{eqnarray} \partial_\tau p_1^{(0)}(\tau) &= 0, \label{eq:pf0}\\
\partial_\tau p_1^{(1)}(\tau) &=
-\left[\tilde{k}_{12}(\tau)+\tilde{k}_{21}(\tau)\right]p_1^{(0)}(\tau)+\tilde{k}_{21}(\tau) \label{eq:pf1}
\end{eqnarray} with $\tilde{k}_{ij}(t)=k_{ij}(t)/k_0$. Equation~\eref{eq:pf0} shows that the zeroth order
is time-independent. Thus, we consider the integral
\begin{eqnarray} &\int_0^\mathcal{T}p_1^{(1)}(t)v_1(t)\dif
t=-\int_0^{2\pi}\partial_\tau p_1^{(1)}(\tau)V_1(\tau)\dif\tau
\nonumber\\
&=\int_0^{2\pi}\left[\left(\tilde{k}_{12}(\tau)+\tilde{k}_{21}(\tau)\right)p_1^{(0)}(\tau)-\tilde{k}_{21}(\tau)\right]V_1(\tau)\dif\tau\stackrel{!}{=}0 \label{eq:int_f}
\end{eqnarray} with $\partial_tV_1(t)=v_1(t)$. Since $p_1^{(0)}(\tau)$ is a constant in time, the expression in square brackets in
the integral~\eref{eq:int_f} is a function of $\Delta
E(\tau)$ only. This suggests the choice $V_1(\tau)=\partial_\tau\Delta E(\tau)/N_1$,
which translates into $v_1(t)=\partial_t^2\Delta E(t)/(N_1\Omega)$ with arbitrary constant $N_1$ with similar
reasoning as previously discussed for slow driving. To summarize, we have found the suitable first two basis functions
\begin{eqnarray} v_1(t)&=\frac{\partial_t^2\Delta E(t)}{N_1\Omega} \;\;\;\;\;\; \text{and} \;\;\;\;\;\;
 w_1(t)&=\frac{\partial_t\Delta E(t)}{N_2}.
\end{eqnarray} 

These basis functions lead to our first main
result. Analogous to equation~\eref{eq:1}, the phase can be identified as
\begin{equation} \label{eq:phase} \Delta\varphi =
\arctan\left(\frac{b_1}{a_1}\right)=\arctan\left(\frac{\langle\partial_tE(t)\rangle\Omega}{\langle\partial_t^2{E}(t)\rangle}\right).
\end{equation} 
Here, we have chosen the constants
\begin{eqnarray} N_1 = \frac{1}{\mathcal{T}}\int_0^\mathcal{T}
\frac{\left(\partial_t^2\Delta E(t)\right)^2}{\Omega^2}\dif t \;\;\;\;\;\;\text{ and }\;\;\;\;\;\;N_2 =
\frac{1}{\mathcal{T}}\int_0^\mathcal{T}\left(\partial_t\Delta E(t)\right)^2\dif t
\end{eqnarray}
such that the coefficients are given by
\begin{equation} \label{eq:coeffa} 
a_1=\frac{1}{\mathcal{T}}\int_0^\mathcal{T}p_1(t)\frac{\partial_t^2\Delta E(t)}{\Omega}\dif t=\langle\partial_t^2{E}(t)\rangle/\Omega
\end{equation}
and \begin{equation}\label{eq:coeffb} 
b_1=\frac{1}{\mathcal{T}}\int_0^\mathcal{T} p_1(t)\partial_t\Delta E(t)\dif t =\langle\partial_tE(t)\rangle.
\end{equation}
The average $\langle\cdot\rangle$ is defined over
many realizations and over one period as
\begin{equation} \langle X_i \rangle \equiv
\sum_{i=1}^N\frac{1}{\mathcal{T}}\int_0^\mathcal{T} p_i(t)X_i(t)\dif t.
\end{equation} 
Furthermore, $b_1$ is the total applied
power~\eref{eq:powers} and, hence, coincides with the total entropy
production rate~\eref{eq:entropy}. We note that the second case of equation~\eref{eq:1} 
can be omitted in equation
\eref{eq:phase} since $a_1$ and $b_1$ are both positive as shown in
\ref{asec:proof_bound}.
Equation~\eref{eq:phase} shows that in the
limit of slow driving $\Delta\varphi$ vanishes due to $b_1=0$. For fast
driving, the phase approaches its maximum with $\Delta\varphi=\pi/2$ due to
$a_1=0$.

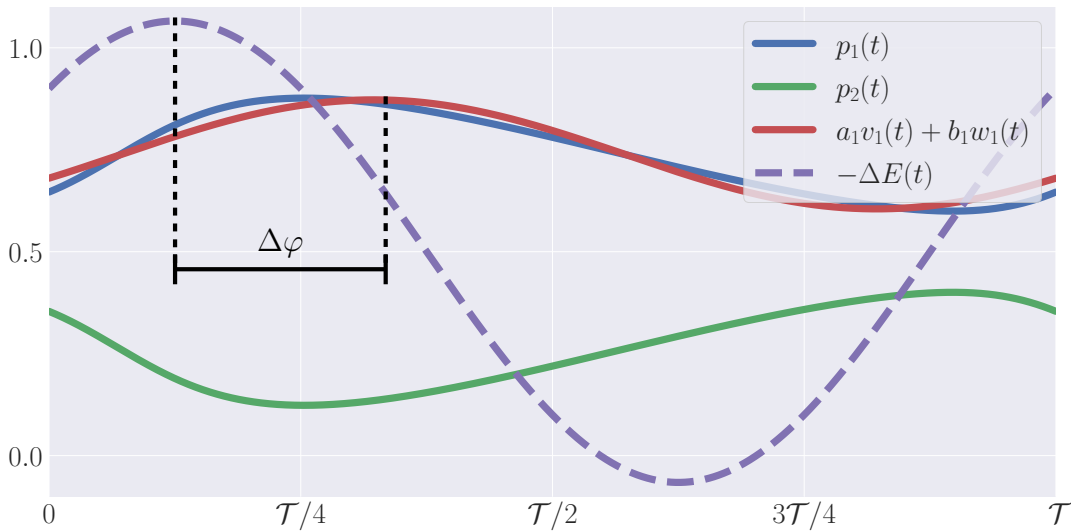
\begin{figure}[t] \centering
  \begin{tikzpicture}[scale=1] 
  \def\a{-4.8}; 
  \def\b{3.35}; 
  \def\c{-2};
  \def\d{2.3}; 
  \def\e{0}; 
  \node at (0,0){\includegraphics[scale=0.34]{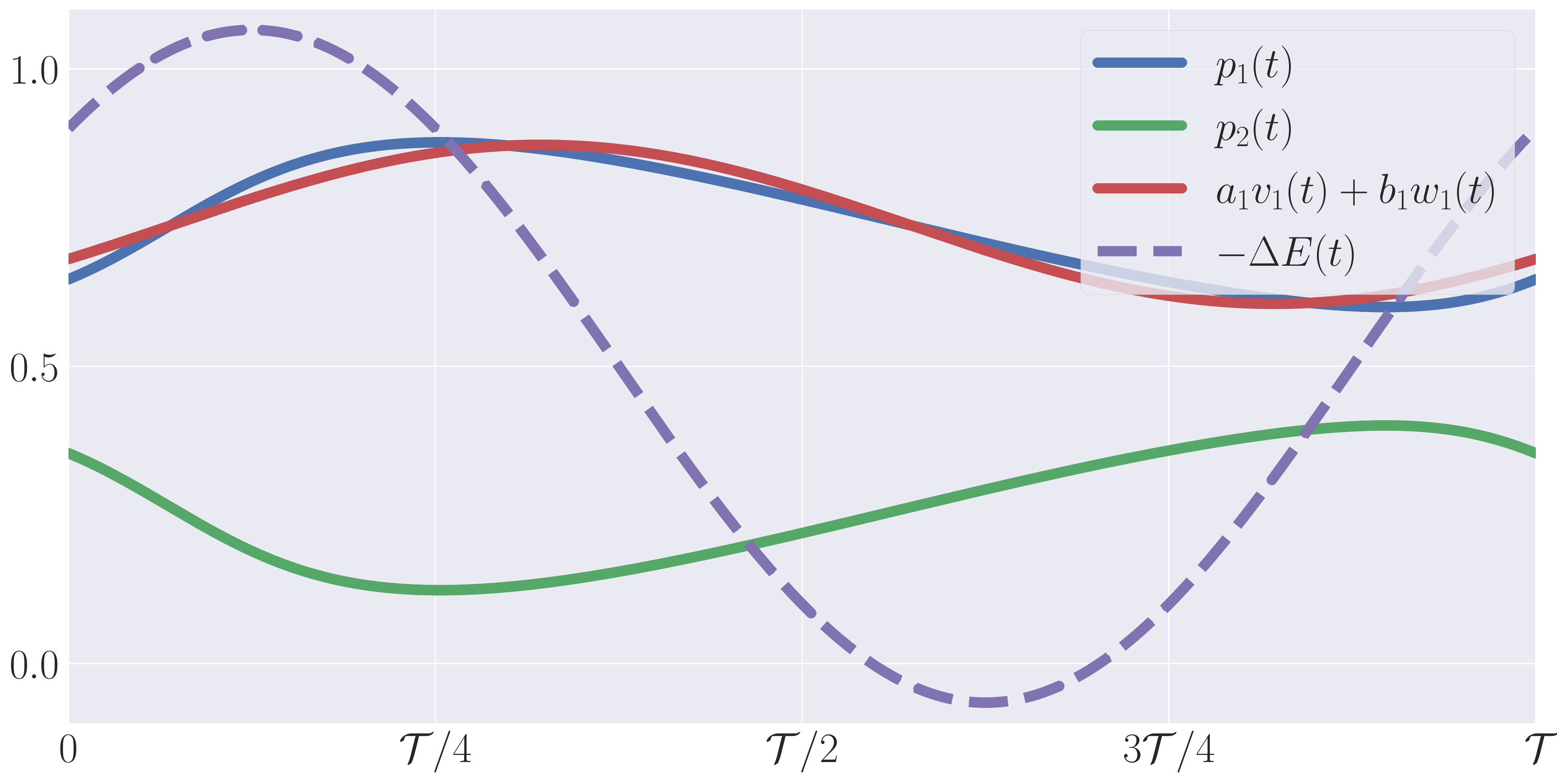}};
  \draw[dashed, black,ultra thick] (\a, \b) -- (\a,\e); 
  \draw[dashed, black,ultra thick](\c, \d) -- (\c, \e); 
  \draw[ultra thick] (\a,\e-0.2) -- (\a,\e+0.2);
  \draw[ultra thick] (\c,\e-0.2) -- (\c,\e+0.2); 
  \draw[ultra thick] (\a,\e) -- node[above, midway]{$\Delta\varphi$} (\c,\e);
  \end{tikzpicture}
  \caption{Illustration of the phase $\Delta\varphi$ for the case of a single
driving frequency~\eref{eq:simplest_case}. The green and blue lines
show the periodic stationary state as a function of time for one period. The
red line shows the first order of the series~\eref{eq:series} shifted along
the $y$-axis by a constant offset. The dashed line indicates the driving
$\Delta E(t)$ scaled along the $y$-axis. The phase $\Delta\varphi$ as defined
in~\eref{eq:phase} is the difference between the extremum of the driving 
and the extremum of $a_1v_1(t)+b_1w_1(t)$.}
  \label{fig:1}
\end{figure}

We now consider the simplest non-trivial case of a single driving frequency to get
a better understanding of the quantity $\Delta\varphi$. The driving can be
written as
\begin{equation} \label{eq:simplest_case} \Delta E(t)=\Delta
E_0+E\sin\left(\Omega t\right)
\end{equation}
with $\Delta E_0\equiv E^0_1-E^0_2$. In this case, the series in equation~\eref{eq:series} reduces to a Fourier
series, since ${v_1(t)=-2\sin\left(\Omega t\right)/(E\Omega)}$ and ${w_1(t)=2\cos\left(\Omega t\right)/(E\Omega)}$. The quantity $\Delta\varphi$ is then the ordinary phase difference between the driving
$\Delta E(t)$ and the probability $p_1(t)$ at the frequency of the driving as
illustrated in Figure~\ref{fig:1}. The example demonstrates the geometric
aspect that is related to the delay in a periodically driven system. In the
more general case of an arbitrary number of driving frequencies the idea of
expressing the probability in terms of the driving remains the same.

Notably, the phase difference does not scale with the amplitude of the driving
$\Delta E(t)$ or the probability $p_1(t)$, neither in the aforementioned
simplest case~\eref{eq:simplest_case}, nor in the general case
\eref{eq:general_case}. However, there is an indirect dependence, because the amplitude
of $\Delta E(t)$ influences the shape of the probability $p_1(t)$ and,
therefore, also the phase. Two functions $A\Delta E(t)$ and
$Bp_1(t)$ yield the same value of $\Delta\varphi$ independently of
the arbitrary constants $A$ and $B$. This is in contrast to the definition of lag
in Refs.~\cite{vaik09, frez17} where small amplitudes inevitably lead to a small
lag. However, the lag does not necessarily vanish for weak driving. As it turns out 
it rather becomes maximal as we will discuss in section \ref{sec:main_result}. We argue that this
phase $\Delta\varphi$, which allows a lag of the order of $1$ even for small driving, is a
meaningful way of quantifying the physical lag.

\begin{figure}[t] \centering \includegraphics[scale=0.35]{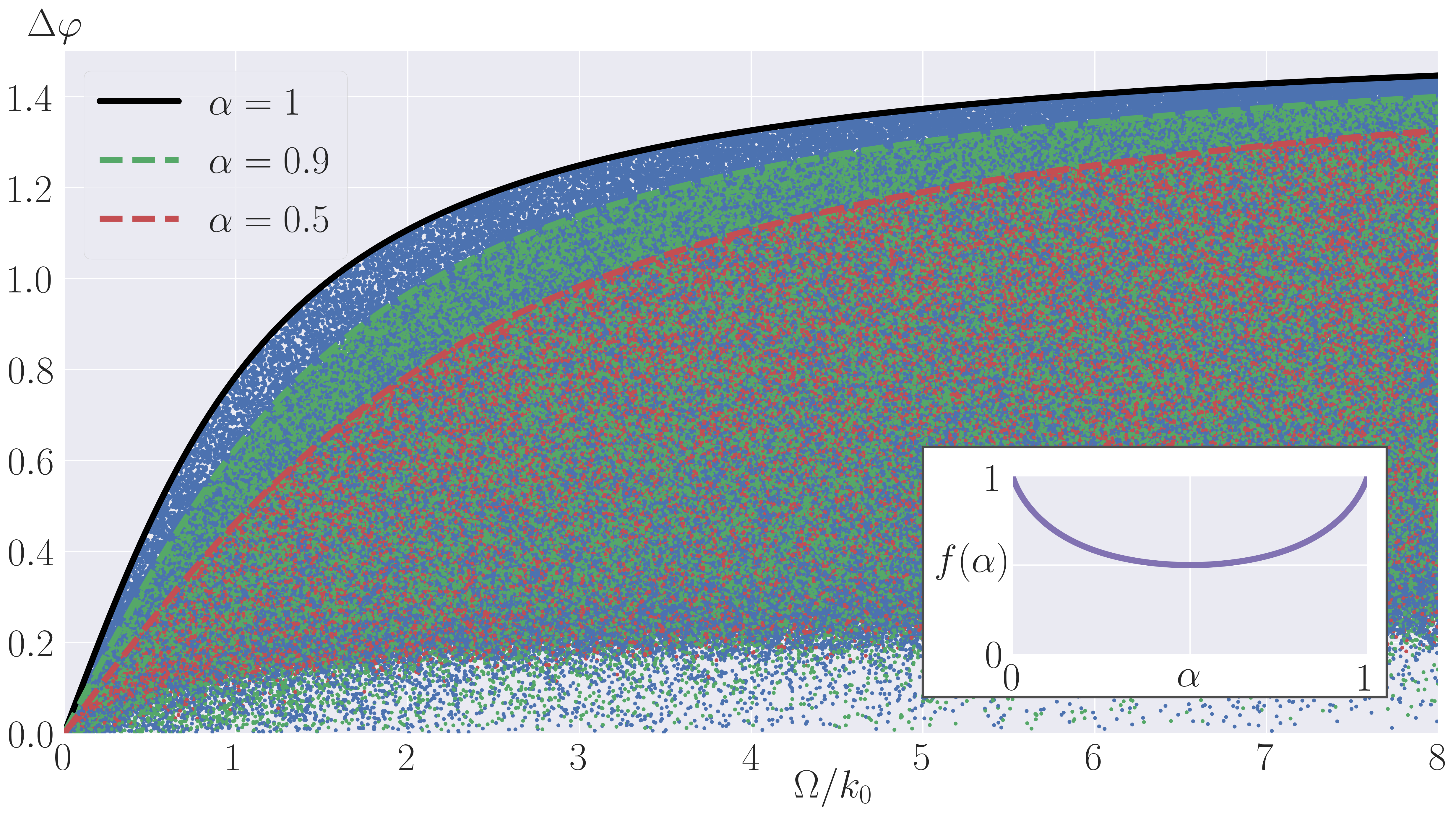}
  \caption{Illustration of the bounds~\eref{eq:bound_1} and
\eref{eq:bound_2}. The colors indicate the value of $\alpha$; blue:
$\alpha\in\left\{0,1\right\}$, green: $\alpha\in\left\{0.1,0.9\right\}$, red:
$\alpha=0.5$. For each system the parameters are drawn randomly from uniform
distributions with $\alpha\in\left\{0,0.1,0.5,0.9,1\right\}$,
$E\in\left[0,4\right]$, $\Omega/k_0\in\left[0.01,8\right]$,
$c_i^n,d_i^n\in\left[-1,1\right]$ and $E_i^0\in\left[-4,4\right]$. The solid black
line shows the bound~\eref{eq:bound_1}. The dashed blue and green lines show
the bound~\eref{eq:bound_2} for $\alpha=0.5$ and $\alpha=0.9$
respectively. The inset shows the $\alpha$ dependency of the bound
\eref{eq:bound_2}.}
  \label{fig:2}
\end{figure}

\section{Bounds on the phase} \label{sec:main_result}
Our second main result is an upper bound on the phase,
\begin{equation} \label{eq:bound_1} \Delta\varphi \leq
\arctan\left(\frac{\Omega}{k_0}\right),
\end{equation} that depends only on the relative speed of the driving and
holds for arbitrary periodic driving~\eref{eq:general_case} in a two-state system. Since $\Omega$
determines the time scale of the driving and $k_0$ the time scale at which
transitions in the system occur, the ratio of these two quantities gives an effective speed of
driving. Figure~\ref{fig:2} illustrates the quality of this bound through
numerical results. Each dot represents the numerically calculated phase
$\Delta\varphi$ in a system with randomly chosen parameters. Obviously, the 
bound~\eref{eq:bound_1} can be saturated for any
relative driving speed $\Omega/k_0$.

The information about the relative speed of the driving alone, encoded in the
parameter $\Omega/k_0$, is sufficient to bound the lag
$\Delta\varphi$. However, by including additional information, specifically about
the distribution of the rates in forward and backward rate as quantified by the
parameter $\alpha$, a stronger bound follows as
\begin{equation} \label{eq:bound_2} \Delta\varphi
\leq\arctan\left(f\left(\alpha\right)\frac{\Omega}{k_0}\right)\leq\arctan\left(\frac{\Omega}{k_0}\right)
\end{equation} with
\begin{equation}
f\left(\alpha\right)\equiv\alpha^\alpha\left(1-\alpha\right)^{1-\alpha}\leq 1
\end{equation}
for all $\alpha\in\left[0,1\right]$.
For any $\alpha$ in the relevant interval $\left[0,1\right]$, the relation
$1/2\leq f\left(\alpha\right)\leq1$ applies. More specifically,
$f\left(\alpha\right)$ is maximized at the boundaries $\alpha=0,1$ and
minimized at the center $\alpha=1/2$ while being monotonous in between and
symmetrical with respect to the center, see inset of Figure~\ref{fig:2}. The
stronger bound is illustrated for a few values of $\alpha$ in Figure
\ref{fig:2}.

For a proof of these bounds, it is sufficient to derive the more detailed bound~\eref{eq:bound_2}, from
which the main result~\eref{eq:bound_1} follows due to
$f\left(\alpha\right)\leq1$, the positivity of the arguments of the arctan-function
and its monotonicity. To derive the bound~\eref{eq:bound_2}, we consider the putative inequality
\begin{equation} \label{eq:proof_start}
\gamma\langle\partial_t^2E(t)\rangle/\Omega-\langle\partial_tE(t)\rangle\geq 0
\end{equation}
with the additional free parameter $\gamma$. Any choice for $\gamma$ that fulfills the inequality \eref{eq:proof_start}
implies a bound of the form
\begin{equation} \label{eq:gamma_bound}
\Delta\varphi\leq\arctan\gamma,
\end{equation}
due to \eref{eq:phase}.
To obtain the best bound, we optimize with respect 
to $\gamma$ while still preserving the inequality \eref{eq:proof_start} as further detailed in
\ref{asec:proof_bound}. This procedure leads to the optimal value of $\gamma$ given by
\begin{equation} 
\gamma^*=f\left(\alpha\right)\frac{\Omega}{k_0},
\end{equation}
which implies the bound~\eref{eq:bound_2} by using~\eref{eq:gamma_bound}.
To summarize, we have derived two bounds on the phase
$\Delta\varphi$, one general bound, which only depends on the relative speed of the driving and one more detailed bound that additionally takes
information about the rate splitting into account.

As the numerical data shown in Figure~\ref{fig:2} demonstrates, the
bound~\eref{eq:bound_2} can be saturated for any relative driving speed
$\Omega/k_0$. We now discuss the conditions leading to
saturation. Saturating the bound is equivalent to constructing a system with
maximum lag. As shown in detail in~\ref{asec:saturation}, the two key
parameters to construct such a system are the amplitude of the driving
$E$ and the energy offset $\Delta E_0$. Remarkably, the exact form of the
driving defined through its Fourier coefficients
\eref{eq:general_case} does not matter in this case. The energy amplitude
needs to be small $E\ll 1$, leading to an overall weak driving in a linear
response regime. The energy offset has to be chosen according to the splitting
of the rates as
\begin{equation} \label{eq:DelE0opt} \Delta E_0 =
\log\left(\frac{1-\alpha}{\alpha}\right).
\end{equation} Qualitatively, both of these conditions are expected for a
system with maximum lag. The system always relaxes towards the instantaneous
equilibrium state but ultimately always lags behind. The speed at which it
relaxes and, therefore, tries to follow the instantaneous equilibrium state is
determined by the transition rates. While the transition rates scale
exponentially with the amplitude $E$ of the driving, the instantaneous
equilibrium state has a weaker scaling, especially since it is limited to
values between $0$ and $1$. Thus, a high energy amplitude generally reduces
the lag or, conversely, a low energy amplitude generally leads to a higher lag.

The energy offset $\Delta E_0$ that leads to the maximum lag depends on the
splitting parameter $\alpha$. The splitting determines which rate is more
sensitive to the energy difference between the states. Choosing $\alpha$ close to $1$
implies that $k_{12}(t)$ is more sensitive than $k_{21}(t)$ and vice versa for $\alpha$ close
to $0$. A positive $\Delta E_0$ increases the rate
$k_{12}(t)$ and decreases the rate $k_{21}(t)$, whereas a negative $\Delta E_0$ has
the opposing effect on the rates. To maximize the lag, the energy offset
$\Delta E_0$ has to increase the rate that is less dependent on the energy
difference and decrease the rate that is more dependent on it. In the extreme 
cases of $\alpha=0,1$ one rate does not scale with
the energy difference at all. Here, $\Delta E_0$ diverges with the expected sign,
as shown by equation~\eref{eq:DelE0opt}.

\section{$N$-state system}
In this section, we consider an arbitrary Markov network with $N$ states.
We aim to generalize the definition
of the phase to quantify the inherent lag in any periodically driven Markov
network. Multiple challenges arise when dealing with these more complex
systems. First, the driving cannot be described by a single scalar quantity
comparable to $\Delta E(t)$. Instead, there are individual energy
differences for all links of the system. Second, there is no single characteristic
time scale of the system. The local detailed
balance relation allows one to have different rate amplitudes $k_{0,ij}$ for every
link. Third, similar to the rate amplitudes, local detailed balance also
allows for a different splitting of rates $\alpha_{ij}$ at any available link.

In the definitions~\eref{eq:coeffa} and~\eref{eq:coeffb} we have written the
coefficients as averages of the second and first derivative of the energies. A natural generalization of the definitions in~\eref{eq:phase},
\eref{eq:coeffa} and~\eref{eq:coeffb} is to keep these notions and apply the
general average
\begin{eqnarray} a_1(N) &=
\frac{\langle\partial_t^2{E}(t)\rangle}{\Omega}=\sum_{i=1}^N\int_0^\mathcal{T}
p_i(t)\frac{\partial_t^2{E}_i(t)}{\Omega}\dif t=\int_0^\mathcal{T}
\mathbf{p}(t)\cdot\partial_t^2{\mathbf{E}}(t)\frac{1}{\Omega}\dif t, \\ b_1(N) &=
\langle\partial_tE(t)\rangle=\sum_{i=1}^N\int_0^\mathcal{T} p_i(t)\partial_tE_i(t)\dif
t=\int_0^\mathcal{T}\mathbf{p}(t)\cdot\partial_t{\mathbf{E}}(t)\dif t
\end{eqnarray} 
with
\begin{equation}
\textbf{p}(t)=\left(p_1(t),\cdots,p_N(t)\right)^T
\end{equation} 
and 
\begin{equation}
\textbf{E}(t)=\left(E_1(t),\cdots,E_N(t)\right)^T
\end{equation}
to obtain a similar expression
\begin{equation} \label{eq:phase_gen} \Delta\varphi =
\arctan\left(\frac{b_1(N)}{a_1(N)}\right)
\end{equation} for the phase. The interpretation of expressing the phase in
terms of the driving still remains valid. However, the driving is now
described by an $N$-component vector instead of a scalar quantity.

\begin{figure} \centering
\begin{minipage}[t]{.3\textwidth}
\includegraphics[scale=0.35]{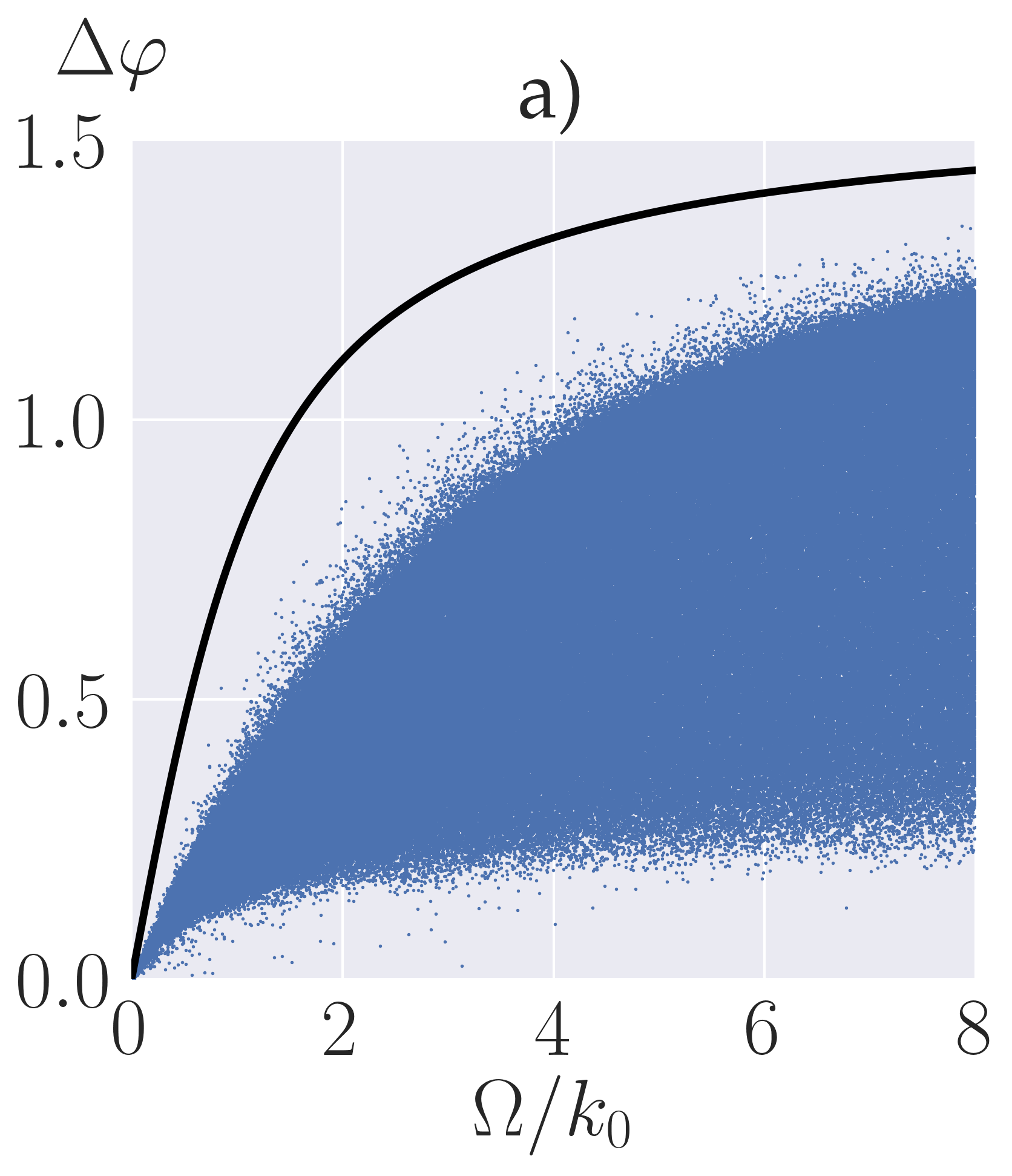}
\end{minipage} \hspace{0.02\textwidth}
\begin{minipage}[t]{.3\textwidth}
\includegraphics[scale=0.35]{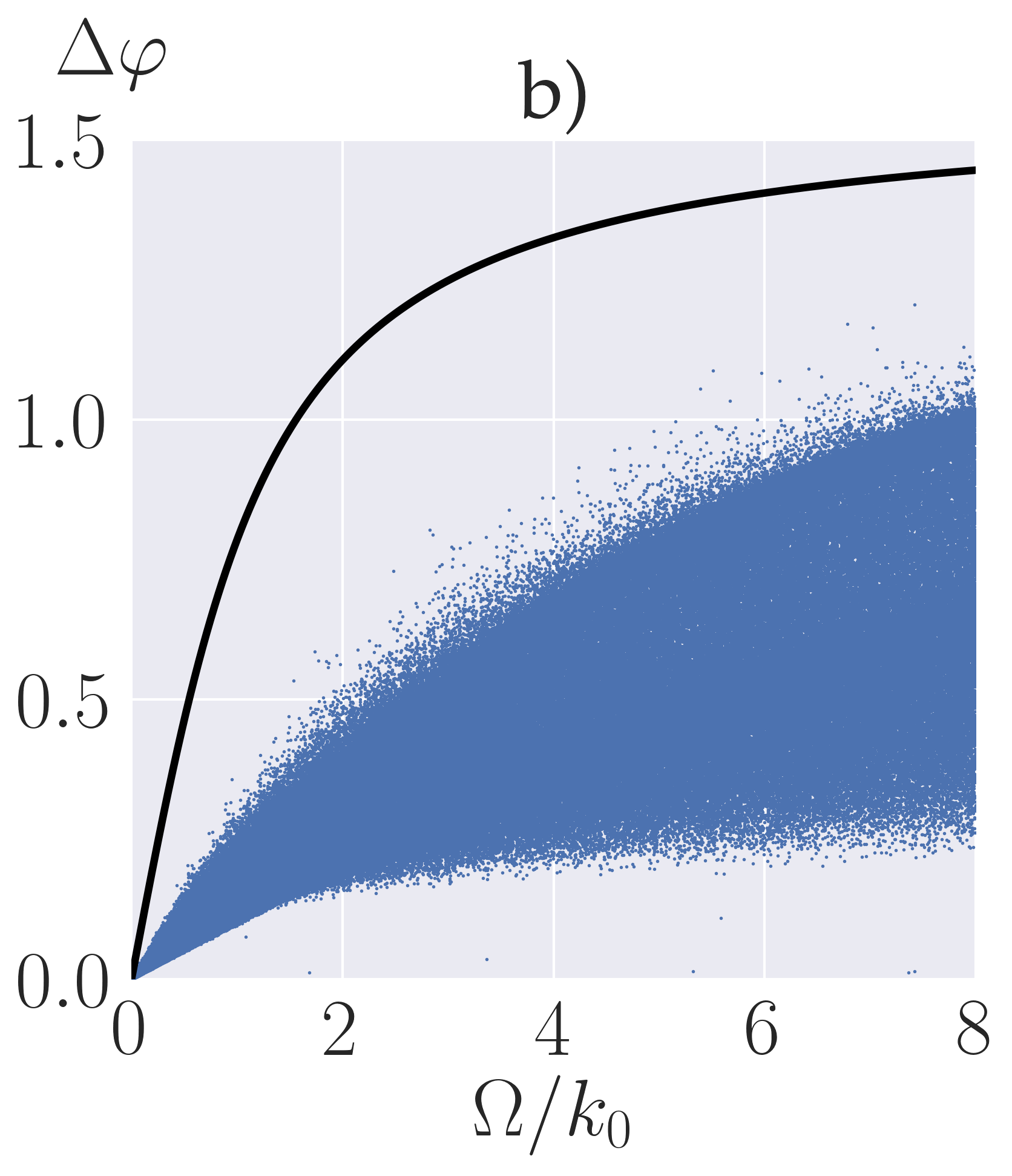}
\end{minipage} \hspace{0.02\textwidth}
\begin{minipage}[t]{.3\textwidth}
\includegraphics[scale=0.35]{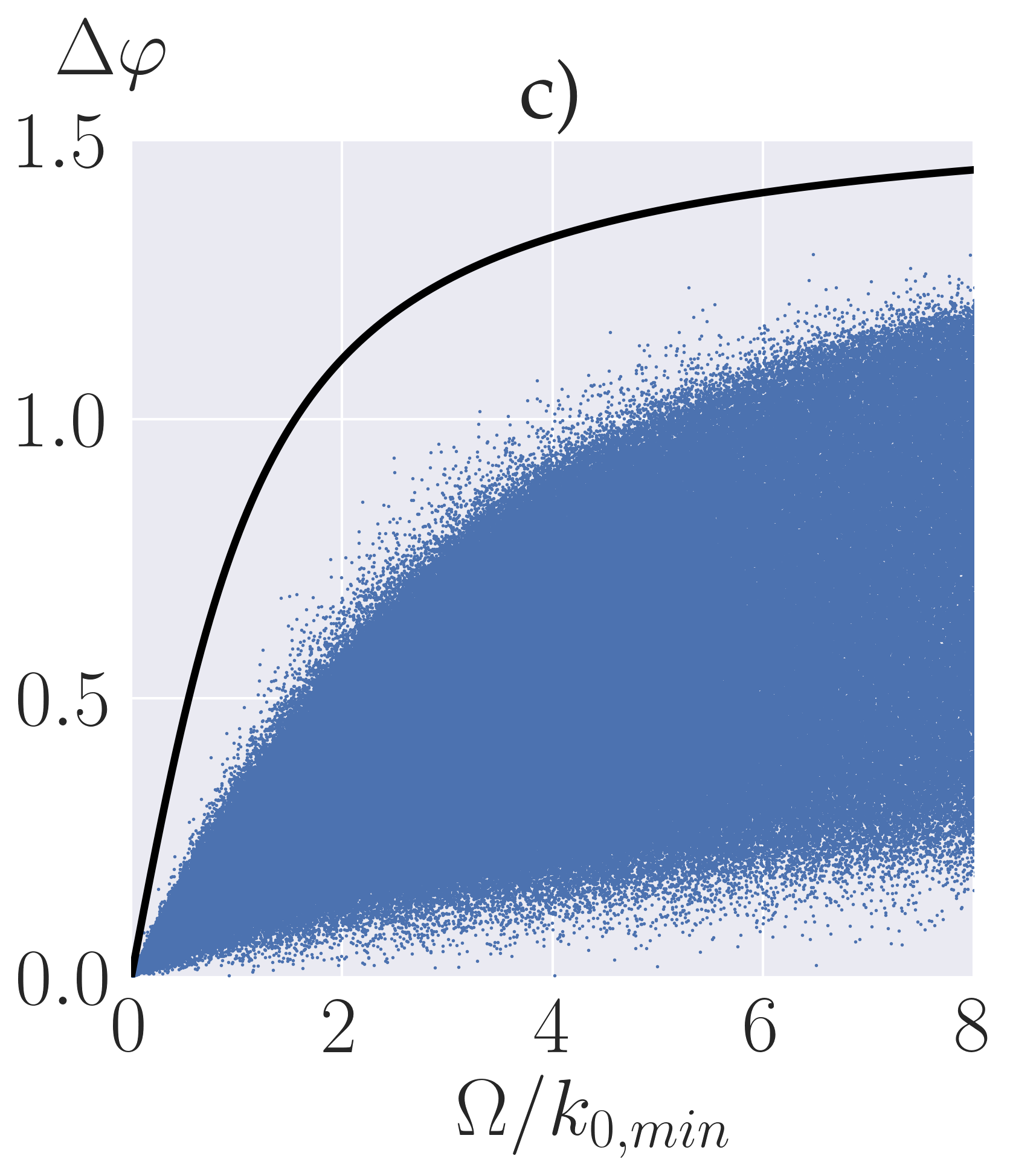}
\end{minipage}
\caption{Illustration of the bound~\eref{eq:bound_1} for the phase
\eref{eq:phase_gen} for $N$-state systems. Each point represents the
numerically calculated phase of a system in dependence of the relative driving
speed. For each system the numerical parameters
$\alpha_{ij}\in\left[0,1\right]$, $E\in\left[0,4\right]$,
$c_i^n,d_i^n\in\left[-1,1\right]$ and
$E_i^0\in\left[-4,4\right]$ are drawn randomly from uniform distributions. a)
$N=3$ state systems with common rate amplitude $k_0$. $\Omega/k_0\in\left[0.01,8\right]$ is
drawn from a uniform distribution. b) $N=5$ and $N=10$ state systems with 
common rate
amplitude $k_0$. $\Omega/k_0\in\left[0.01,8\right]$ is drawn from a uniform
distribution. c) $N=3$ state system with inhomogeneous rate amplitudes $k_{0,ij}\in\left[0.01,8\right]$. The relative driving speed is the ratio of the
frequency $\Omega$ and the minimal rate amplitude
$k_{0,\text{min}}=\min_{i,j}k_{0,ij}$. The distribution for $k_{0,ij}$ was modified
to more closely resemble a uniform distribution in $1/k_{0,\text{min}}$.}
\label{fig:3}
\end{figure}

We now analyze the bound~\eref{eq:bound_1} in the context of such $N$-state
systems. We consider the more general version which allows us to disregard
the rate-splitting $\alpha_{ij}$. In the simplest case, we assume a common
rate amplitude $k_{0,ij}=k_0$ for every available link which ensures a common
time scale. The bound~\eref{eq:bound_1} seems
to hold in this case as well. Although a proof remains to be found,
we demonstrate numerical evidence for $N=3$ state systems in Figure
\ref{fig:3}~a) and for $N=5$ and $N=10$ state systems in Figure~\ref{fig:3}~b) with random
parameters. The analysis of systems with other numbers
of states $N$ yields similar results. For systems with many states, the majority of 
the numerical data is farther from the bound. However, we can
always create an effective $(N-1)$-state system by shifting the time-independent
part of the energy of one state towards infinity. Thus, any $N$-state
system can accomplish the same saturation as a two-state system through
an appropriate choice of parameters.

In a more general case, every link has its own rate amplitude $k_{0,ij}$ and,
therefore, also its own time scale on which transitions along that specific
link $(i,j)$ occur. To establish a bound similar to~\eref{eq:bound_1}, we need
a single parameter representing the time scale of the transitions. We choose
the minimal rate amplitude $k_{0,\text{min}}\equiv\min_{i,j}k_{0,ij}$ since the bound
\eref{eq:bound_1} is an upper bound. We thus conjecture
\begin{equation} \label{eq:gen_bound}
\Delta\varphi\leq\arctan\left(\frac{\Omega}{k_{0,\text{min}}}\right).
\end{equation} The speed of transitions is approximated by the slowest link in
the network. Numerical evidence suggests~\eref{eq:gen_bound} to hold as
demonstrated with examples in Figure~\ref{fig:3} c). Note that a missing link
$(i,j)$ means that the corresponding rates $k_{ij}(t)$, $k_{ji}(t)$ and, therefore,
the rate amplitude $k_{0,ij}$ vanishes. Thus, the bound~\eref{eq:gen_bound}
becomes trivial. A missing link leads to an effective barrier between
areas of the network by shaping the energy landscape accordingly. These
barriers influence and increase the lag additionally which is not captured by
the rate amplitudes alone. Thus, it is not surprising that the bound
\eref{eq:gen_bound} becomes trivial in this case. 

We trust that the phase $\Delta\varphi$ as defined in \eref{eq:phase_gen} remains a suitable quantity to capture
the lag even in this most general case. However, any non-trivial bound for
this quantity requires information about the topology of the network. One possible 
approach would be to establish effective transition rates between any two states, 
even if they do not share a direct link. This would allow one to account for barriers or other 
topological aspects that additionally influence the lag in the system. The challenge 
would be to find the correct way to compute these effective rates.

\section{Conclusion}
In this paper, we have introduced a phase shift that allows one to quantify
the lag for periodic steady states. The phase shift describes the phase
difference between the driving and the corresponding response of the
system. As such, it quantifies less the overall similarity between the driving
and the response but rather focuses specifically on the lag in the
system along the time-axis. For two-state systems, we have analyzed and proven
two upper bounds on this phase. The first bound requires only the relative
driving speed of the system as input. The second bound is stronger since it
additionally takes the splitting in forward and backward rate into
account. Both bounds are tight and can be saturated as we have illustrated, first, by
numerical examples and second, by constructing a family of special linear
response cases that always saturate the bounds. For systems with an arbitrary
number of states we conjecture a bound similar to the general bound in the
two-state case based on numerical evidence. However, a
rigorous proof remains to be achieved since the essential steps of our proof
for two-state systems cannot be applied to systems with an arbitrary number of
states in a straightforward way. Last but not least, a more detailed bound
taking into account information about the topology of the network remains 
to be found for systems with missing links between states.

\newpage
\appendix

\section{Derivation of the bound~\eref{eq:bound_2}}
\label{asec:proof_bound}
We write equation~\eref{eq:proof_start} as
\begin{equation} \label{eq:to_proof}
\int_0^\mathcal{T}\frac{\gamma}{\Omega}(\partial_t^2\Delta E(t))p_1(t)-(\partial_t\Delta E(t))p_1(t)\dif
t\geq0.
\end{equation} 
We note that the second term of the integral is the total entropy production rate as defined in \eref{eq:entropy}. Due to the second law of thermodynamics, proving the inequality \eref{eq:to_proof} for positive $\gamma$ implies $\langle\partial_t^2{E}(t)\rangle/\Omega\geq 0$.

The first step to proof inequality~\eref{eq:to_proof} is to
find the extremal point of the left hand side with arbitrary but fixed driving
$\Delta E(t)$ by calculating the variations with respect to the probabilities and the rates. We require the probabilities and rates to fulfill the master equation \eref{eq:me} and, hence, we introduce a Lagrange multiplier
$\lambda(t)$ leading to the Lagrange functional
\begin{eqnarray}
G[k_{12},p_1]&\equiv\int_0^\mathcal{T}\left[\frac{\gamma}{\Omega}(\partial_t^2\Delta E(t))-(\partial_t\Delta E(t))\right]p_1(t) \label{eq:functional}
\\ &+\lambda(t)\biggl[\partial_t{p}_1(t)+k_{12}(t)\Bigl\{\left(1+e^{-\Delta
E(t)}\right)p_1(t)-e^{-\Delta E(t)}\Bigr\}\biggr]\dif t, \nonumber
\end{eqnarray} where we already inserted the normalization conditions for the
probabilities and the local detailed balance condition to eliminate the
rate $k_{21}(t)$. By calculating the functional derivatives with respect to $p_1(t)$, $k_{12}(t)$ and
$\lambda(t)$ we obtain the equations
\begin{eqnarray}&\frac{\delta G\left[k_{12},p_1\right]}{\delta
k_{12}(t)}=\lambda(t)\Bigl[\left(1+e^{-\Delta E(t)}\right)p_1(t)-e^{-\Delta
E(t)}\Bigr]= 0, \label{eq:fd_con1} \\ 
&\frac{\delta G\left[k_{12},p_1\right]}{\delta
p_1(t)}=\frac{\gamma}{\Omega}\partial_t^2\Delta E(t)-\partial_t\Delta E(t)-\partial_t{\lambda}(t) \label{eq:fd_con2} 
\\ &\quad\quad\quad\quad\quad\quad+\lambda(t)k_{12}(t)\left(1+e^{-\Delta
E(t)}\right)= 0\nonumber
\end{eqnarray} and the master equation as third condition. Equation~\eref{eq:fd_con1} can be solved either by choosing
\begin{equation} p_1(t)=\frac{1}{1+e^{\Delta E(t)}}=p_1^{\text{qs}}(t) \;\;\;\;
\text{or} \;\;\;\; \lambda(t)=0.
\end{equation} The latter solution, $\lambda(t)=0$, fails to satisfy the condition~\eref{eq:fd_con2}. In contrast, when choosing $p_1(t)=p_1^{\text{qs}}(t)$ the only requirement for
$\lambda(t)$ is to solve the ordinary first order differential equation~\eref{eq:fd_con2}. A solution can formally be obtained by variation of constants 
with the boundary condition $\lambda(0)=\lambda(\mathcal{T})$. The probability
$p_1^{\text{qs}}(t)$ is the solution to the master equation in the limit
of slow driving. Thus, the only extremal point of the left hand side of~\eref{eq:to_proof} is obtained in the limit of slow driving $\Omega/k_0\to0$.

The second step of the proof is to show that the functional~\eref{eq:functional} has a minimum for $\Omega/k_0\to0$. Since the extremum is reached for $\Omega/k_0\to0$ independently
of all other parameters, we use a perturbation theory around the
slow driving limit. Rewriting~\eref{eq:to_proof} with a dimensionless time variable
$\tau\equiv\Omega t$ leads to
\begin{equation} \label{eq:I} I\equiv\int_0^{2\pi}\gamma(\partial_\tau^2\Delta
E(\tau))p_1(\tau)-(\partial_\tau\Delta E(\tau))p_1(\tau)\dif \tau.
\end{equation} The perturbation series
\begin{equation} \label{eq:pert_series}
p_1(\tau)=p_1^{\text{qs}}(\tau)+\frac{\Omega}{k_0}p_1^{(1)}(\tau)+\mathcal{O}\left(\frac{\Omega^2}{k_0^2}\right)
\end{equation} shows that the leading order of the second term in
$I$ vanishes in the limit $\Omega/k_0\to0$. The first order of $p_1(t)$
in $\Omega/k_0$ can be calculated by inserting the perturbation series~\eref{eq:pert_series} into
the master equation and is thus given by
\begin{equation} p_1^{(1)}(\tau)=p_1^{\text{qs}}(\tau)-\frac{\partial_\tau
p_1^{\text{qs}}(\tau)}{\tilde{k}_{12}(\tau)+\tilde{k}_{21}(\tau)} \label{eg:first_order_pert_series}
\end{equation} with $\tilde{k}_{ij}(t)=k_{ij}(t)/k_0$. Using~\eref{eg:first_order_pert_series}, 
expanding~\eref{eq:I} up to first order in $\Omega/k_0$ and integrating the first
term by parts yields
\begin{eqnarray} I=\int_0^{2\pi}\dif\tau\frac{\left(\partial_\tau\Delta
E(\tau)\right)^2\exp\left(\Delta E(\tau)\right)}{\left[1+\exp\left(\Delta
E(\tau)\right)\right]^2}\left[\gamma-\frac{\Omega}{k_0}\frac{1}{\tilde{k}_{12}(\tau)+\tilde{k}_{21}(\tau)}\right] \label{eq:int_be_pos}\\
\quad\quad+\mathcal{O}\left(\frac{\Omega^2}{k_0^2}\right). \nonumber
\end{eqnarray} The functional~\eref{eq:functional} has a minimum for $\Omega/k_0\to0$ if the leading order of~\eref{eq:int_be_pos} is positive. The first term of the integrand is always positive. To make the
entire integrand and, thus, also the integral $I$ positive we require the free parameter
$\gamma$ to fulfill the relation
\begin{equation}\label{eq:gamma_req}
\gamma\geq\frac{\Omega}{k_0}\frac{1}{\tilde{k}_{12}(\tau)+\tilde{k}_{21}(\tau)}
\end{equation} for all $\tau$. This still yields the optimal $\gamma$. For the
integral \eref{eq:int_be_pos} to be positive the integrand does not have to be positive
everywhere. However, we can always construct systems that lead
to $I<0$ if $\gamma$ does not fulfill the condition~\eref{eq:gamma_req} by choosing $\Delta E(t)$ accordingly as
we show in~\ref{asec:saturation}. Assuming arbitrary values of $\Delta E(\tau)$ we calculate the maximum of the right hand side of~\eref{eq:gamma_req}. The condition
\begin{equation} \frac{\dif}{\dif\Delta
E(\tau)}\left[\frac{\Omega}{k_0}\frac{1}{\tilde{k}_{12}(\tau)+\tilde{k}_{21}(\tau)}\right]_{\Delta
E(\tau)=\Delta E^*}=0
\end{equation} 
leads to the maximum
\begin{eqnarray}
&\frac{\Omega}{k_0}\frac{1}{\tilde{k}_{12}(\tau)+\tilde{k}_{21}(\tau)}\leq\frac{\Omega}{k_0}\left.\frac{1}{\tilde{k}_{12}(\tau)+\tilde{k}_{21}(\tau)}\right|_{\Delta
E(\tau)=\Delta E^*}\\ 
&=\frac{\Omega}{k_0}\alpha^\alpha\left(1-\alpha\right)^{1-\alpha}\equiv\gamma^*
\end{eqnarray} with $\Delta E^*=\log\left((1-\alpha)/\alpha\right)$ independently of $\tau$. The parameter $\gamma^*$ is the optimal
choice that still leads to a minimum of $I$ in the limit
$\Omega/k_0\to 0$. To summarize, we have
derived the optimal bound~\eref{eq:bound_2} on $\Delta\varphi$ and, thus, our second main result~\eref{eq:bound_1}.

\section{Saturating the bound~\eref{eq:bound_2}}
\label{asec:saturation} In this section, we calculate the phase $\Delta\varphi$ for specific 
parameters that lead to the saturation of the bound~\eref{eq:bound_2}. We consider systems with small energy amplitudes $E\ll 1$ in a linear response regime around equilibrium and calculate the periodic steady state by using perturbation theory
for small $E$ to obtain an expression for the phase. We
consider a driving of the form
\begin{equation} \Delta E(t) = \Delta E_0+E\Delta\tilde{E}(t)
\end{equation} where $\Delta\tilde{E}(t)\equiv\sum_{n=1}^\infty\left[c_n\sin\left(\Omega
t\right)+d_n\cos\left(\Omega t\right)\right]$ is the time-dependent part of the
driving. We expand the rates up to linear order in $E$
\begin{eqnarray} k_{12}(t)&=k_0e^{\alpha\Delta E_0}\left(1+\alpha
E\Delta\tilde{E}(t)\right)+\mathcal{O}\left(E^2\right), \\
k_{21}(t)&=k_0e^{-(1-\alpha)\Delta E_0}\left(1-\left(1-\alpha\right)
E\Delta\tilde{E}(t)\right)+\mathcal{O}\left(E^2\right)
\end{eqnarray} and use the ansatz
\begin{equation} p_1(t)=p_1^{(0)}(t)+Ep_1^{(1)}(t)+\mathcal{O}\left(E^2\right)
\end{equation} for the probability. We plug these into the master equation
\eref{eq:me} together with the normalization condition $p_1(t)+p_2(t)=1$. In zeroth order we obtain the equation
\begin{equation} \partial_tp_1^{(0)}(t)=-\left[k_0e^{\alpha\Delta
E_0}+k_0e^{-(1-\alpha)\Delta E_0}\right]p_1^{(0)}(t)+e^{-(1-\alpha)\Delta E_0}.
\end{equation} The solution is the equilibrium distribution for $E=0$ and reads
\begin{equation} p_1^{(0)}(t)=\frac{1}{1+e^{\Delta E_0}},
\end{equation} where we additionally used the condition
$p_1^{(0)}(0)=p_1^{(0)}(\mathcal{T})$ for the periodic steady state. To first order in $E$ we
obtain
\begin{eqnarray}
p_1^{(1)}(t)=\sum_{n=1}^\infty\frac{-B}{A^2+n^2\Omega^2}&\Bigl[\left(c_nA-d_nn\Omega\right)\sin\left(n\Omega
t\right) \\ &+\left(c_nn\Omega+d_nA\right)\cos\left(n\Omega t\right)\Bigr]
\nonumber
\end{eqnarray} with the constants
\begin{eqnarray} A &= -\left[k_0e^{\alpha\Delta E_0}+k_0e^{-(1-\alpha)\Delta
E_0}\right], \\ B &= -k_0e^{-\left(1-\alpha\right)\Delta
E_0}\left[1-\alpha+\left(\alpha-1+\alpha e^{\Delta
E_0}p_1^{(0)}\right)\right].
\end{eqnarray} Since the zeroth order is time-independent the phase
$\Delta\varphi$ is determined by the first order of $p_1(t)$. Calculating the
coefficients $a_1$ and $b_1$ from the definition of the phase~\eref{eq:phase}
according to equations~\eref{eq:coeffa} and~\eref{eq:coeffb} leads to
\begin{eqnarray} \langle\partial_tE(t)\rangle &=
\sum_{n=1}^\infty\frac{-B}{A^2+n^2\Omega^2}n^2\Omega\pi\left(c_n^2+d_n^2\right),\\
\langle\partial_t^2{E}(t)/\Omega\rangle &=
\sum_{n=1}^\infty\frac{B}{A^2+n^2\Omega^2}An^2\pi\left(c_n^2+d_n^2\right)
\end{eqnarray} and, thus,
\begin{equation} \label{eq:lr_phase}
\Delta\varphi_\text{lr}=\arctan\left(\frac{\Omega}{-A}\right)
\end{equation} for the phase in the linear response regime.

Next, we optimize all parameters in the expression~\eref{eq:lr_phase} for the phase
that are not present in the bound~\eref{eq:bound_2}. Through this procedure,
we obtain the maximum phase $\Delta\varphi$ in the linear response
regime. In this case the only remaining parameter is the energy offset
$\Delta E_0$. The condition
\begin{equation} \left.\frac{\dif}{\dif \Delta
E_0}\arctan\left(\frac{\Omega}{-A}\right)\right|_{\Delta E_0=\Delta
E_0^*}=0
\end{equation} leads to the optimal energy offset given by
\begin{equation} \Delta E_0^*=\log\left(\frac{1-\alpha}{\alpha}\right).
\end{equation} Inserting this optimal energy offset into the expression for
the phase in the linear response regime~\eref{eq:lr_phase} yields the
maximal phase
\begin{equation}
\Delta\varphi_{\text{lr},\text{max}}=\arctan\left(\frac{\Omega}{k_0}\alpha^\alpha\left(1-\alpha\right)^{1-\alpha}\right),
\end{equation} which is exactly the right hand side of the bound
\eref{eq:bound_2}. This also implies that the more general bound
\eref{eq:bound_1} can be saturated in linear response for
systems where the parameter for the rate splitting is additionally restricted to $\alpha=0,1$.

\section*{References}
\bibliographystyle{iopart-num}
\bibliography{./Bibliography/refs.bib}
\end{document}